\documentclass{jps-cp}
\usepackage{txfonts} 
\usepackage{amsmath,amssymb,bm,graphicx,bbold,epsf,colordvi}
\usepackage{slashed}
\def\bea#1\eea{\begin{align}#1\end{align}} 
\usepackage{lipsum}
\usepackage{soul}
\usepackage{braket}
\allowdisplaybreaks 
\usepackage{color}
\usepackage[dvipsnames]{xcolor}

\usepackage{mathrsfs}
\usepackage{appendix}
\usepackage{lipsum}

\newcommand{\bef}{\begin{figure}[hbt]\centering}
\newcommand{\eef}{\end{figure}}

\usepackage{mathtools}
\usepackage{subfigure}
\usepackage{booktabs}
\usepackage{graphicx}
\graphicspath{ {./figure/} }
\usepackage{lipsum}

\newcommand{\beq}{\begin{equation}}
\newcommand{\eeq}{\end{equation}}
\def\bea#1\eea{\begin{align}#1\end{align}}

\def \be {\begin{equation}}
\def \ee {\end{equation}}
\def \ba {\begin{eqnarray}}
\def \ea {\end{eqnarray}}

\allowdisplaybreaks

\title{Spin Asymmetries in Electron-jet Production at the EIC}

\author{Zhong-Bo \textsc{Kang}$^{1,2,3}$, Kyle \textsc{Lee}$^{4,5}$, Ding Yu \textsc{Shao}$^{6,7}$ and Fanyi \textsc{Zhao}$^{1,2}$}

\inst{$^{1}$Department of Physics and Astronomy, University of California, Los Angeles, CA 90095, USA \\
$^{2}$Mani L. Bhaumik Institute for Theoretical Physics, University of California, Los Angeles, CA 90095, USA \\
$^{3}$Center for Frontiers in Nuclear Science, Stony Brook University, Stony Brook, NY 11794, USA \\
$^{4}$Nuclear Science Division, Lawrence Berkeley National Laboratory, Berkeley, CA 94720, USA \\
$^{5}$Physics Department, University of California, Berkeley, CA 94720, USA \\
$^{6}$Department of Physics and Center for Field Theory and Particle Physics, Fudan University, Shanghai 200433, China \\
$^{7}$Key Laboratory of Nuclear Physics and Ion-beam Application (MOE), Fudan University, Shanghai 200433, China
}

\email{zkang@ucla.edu, kylelee@lbl.gov, dingyu.shao@cern.ch, fanyizhao@physics.ucla.edu}

\recdate{January 12, 2022}

\abst{We investigate all the possible spin asymmetries that can occur in back-to-back electron-jet production with hadron observed inside a jet in electron-proton collisions. We derive the factorization formalism for all spin asymmetries and perform phenomenological studies for the future electron ion collider kinematics. We illustrate that the back-to-back electron-jet production opens up new opportunities to study transverse momentum dependent fragmentation functions and distribution functions.}

\kword{transverse momentum structure (TMD), spin asymmetry, perturbative QCD, jet physics}

\begin{document}
\maketitle

\section{Introduction}
The internal structure of jets has been an active research topic in QCD in recent years. Back-to-back electron+jet creation and the related jet substructure have lately been presented as innovative probes of both the transverse momentum dependent parton distribution functions (TMDPDFs) and the transverse momentum dependent fragmentation functions at the EIC~\cite{Liu:2018trl,Arratia:2020nxw,Liu:2020dct}. In this work, we present the general theoretical framework for the hadron distribution in a jet in back-to-back electron-jet production from electron-proton~($ep$) colliders, where both incoming particles and outgoing hadrons inside the jet have general polarizations. We measure ${\boldsymbol q}_T$, the imbalance of the transverse momentum of the final-state electron and the jet, $z_h$, the momentum fraction of the jet carried by the hadron, and ${\boldsymbol j}_\perp$, the transverse momentum of the hadron inside the jet relative to the jet axis. As shown below, we can constrain the TMDPDFs and TMDFFs independently applying the simultaneous differential information on ${\boldsymbol q}_T$ and ${\boldsymbol j}_\perp$. Specifically, ${\boldsymbol j}_\perp$ is exclusively sensitive to TMDFFs while the ${\boldsymbol q}_T$-dependence can only be affected by the TMDPDFs. Compared to the semi-inclusive deep inelastic scattering (SIDIS) process without jet production~\cite{Bacchetta:2006tn}, where TMDFFs and TMDPDFs are always convolved, auxiliary processes like dihadron production in electron-positron annihilations and Drell-Yan production in proton-proton collisions are needed for constraining TMDFFs and TMDPDFs separately.

\section{Theoretical framework}\label{sec:th}
\begin{figure}
  \centering
  \includegraphics[width=3.2in,trim={0.3cm 0.3cm 0.3cm 0.3cm},clip]{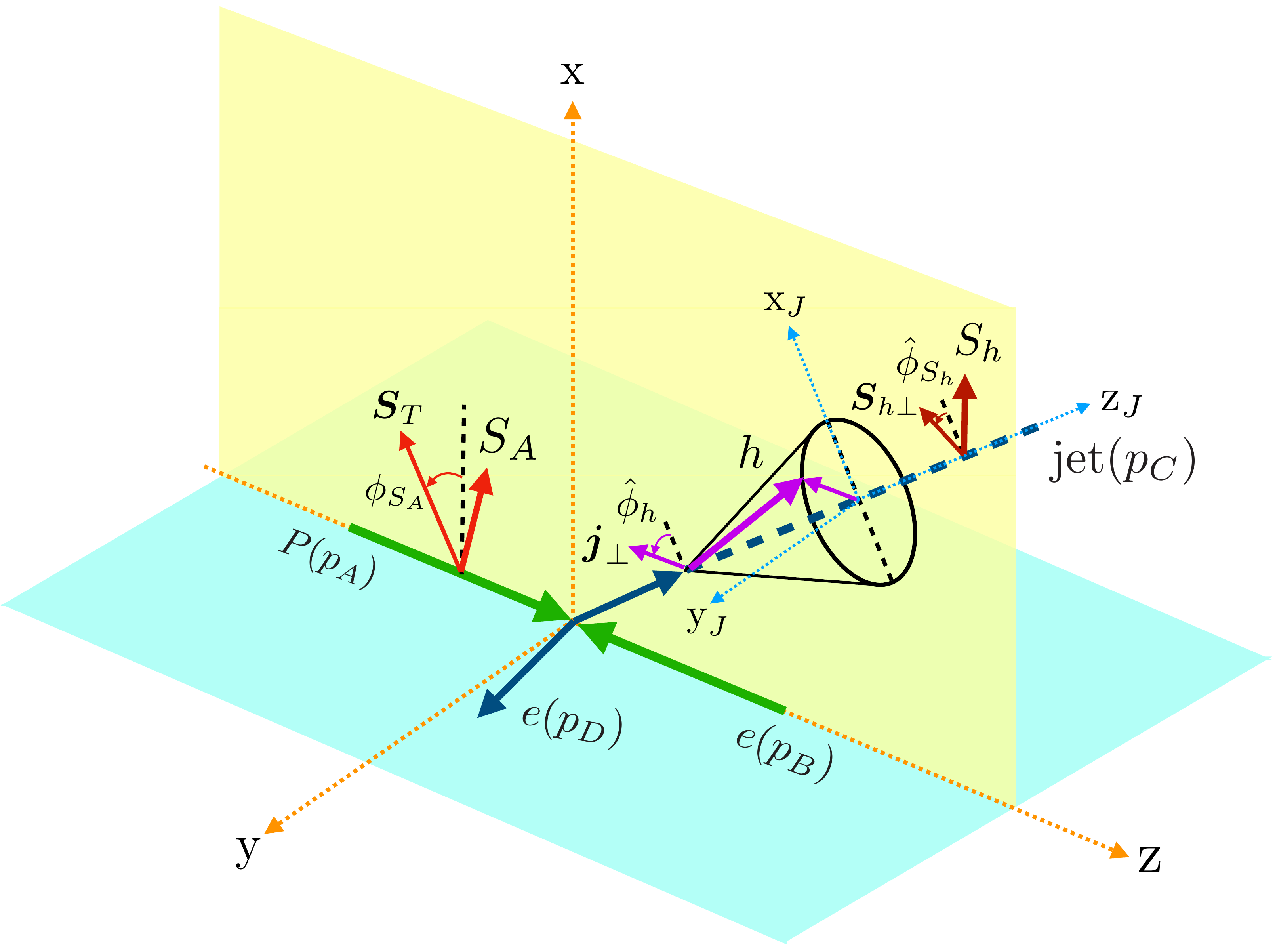} 
  \caption{Hadron in jet with electron production from back-to-back $ep$ collision, where $S_A$ is the spin of the incoming proton, $S_{h}$ represents the spin of the produced hadron in jet. The colliding direction together with the jet axis defines the $xz$-plane.}
  \label{fig:jperp}
\end{figure}

In this section, we provide the theoretical framework of the back-to-back electron-jet production, where the distribution of hadrons are measured inside the jet, 
\bea
p({p}_A,{S}_{A})+e({p}_B, \lambda_e)\rightarrow \Big[\text{jet}({p}_C)\, h\left(z_h, \bm{j}_\perp,\bm{S}_{h\perp}\right)\Big]+e({p}_D)+X\,,
\eea
where $z_h$ is the longitudinal momentum fraction of the jet carried by the hadron $h$, $\bm{j}_\perp$ and $\bm{S}_{h\perp}$ are the hadron transverse momentum and transverse spin with respect to the jet axis. This scattering is comprehensively illustrated in Fig.~\ref{fig:jperp}. We parametrize the transverse momentum imbalance $\bm{q}_T \equiv \bm{p}_{CT} + \bm{p}_{DT}$ and transverse spin vector $\bm{S}_T$  of the incoming proton in terms of their azimuthal angles as
\bea
\label{eq:sT}
\bm{q}_T = q_T(\cos\phi_{q},\sin\phi_{q})\,,\quad \bm{S}_T = S_T(\cos\phi_{S_A},\sin\phi_{S_A})\,,
\eea
where the subscript $T$ indicates that the vector is transverse to the beam direction. On the other hand, we use $\perp$ to denote the vectors transverse to the jet axis for the jet helicity frame. We parametrize the unit vector of transverse momentum ${\bm{j}}_\perp$ and the transverse spin ${\bm S}_{h\perp}$ in the $ep$ center-of-mass frame as
\bea
\hat{\bm{j}}_{\perp}=(\cos\hat{\phi}_h\cos\theta_J,\sin\hat{\phi}_h,-\cos\hat{\phi}_h\sin\theta_J)\,,\quad\hat{\bm{S}}_{h\perp}=(\cos\hat{\phi}_{S_h}\cos\theta_J,\sin\hat{\phi}_{S_h},-\cos\hat{\phi}_{S_h}\sin\theta_J),
\label{eq:bmj}
\eea 
where $\theta_J$ represents the polar angle of the jet momentum $\bm{p}_C$ measured relative to the beam direction and $\hat{\phi}_{h}$ is the azimuthal angle of the produced hadron transverse momentum $\bm{j}_\perp$ in the jet frame $x_J y_J z_J$ shown in Fig.~\ref{fig:jperp}. Then the differential cross section is given by 
\bea
\label{eq:unpjeth}
&\frac{d\sigma^{p(S_A)+e(\lambda_e)\to e+(\text{jet}\,h)+X}}{d{p}^2_Tdy_Jd^2{\bm q}_Tdz_h d^2{\bm j}_\perp}=F_{UU,U}+\cos(\phi_{q}-\hat{\phi}_h)F^{\cos(\phi_{q}-\hat{\phi}_h)}_{UU,U}+\cdots
\eea
where $F_{AB,C}$ is the spin-dependent structure functions with $A$, $B$, and $C$ representing the polarization of the incoming proton, electron, and the outgoing hadron inside the jet, respectively. (The full expression is provided in~\cite{Kang:2021ffh}.) 
\section{Example 1: Unpolarized $\pi^\pm$ in jet}\label{sec:un_h}
In this section, we study the azimuthal asymmetry $\cos(\phi_{q}-\hat{\phi}_h)$ in Eq.~\eqref{eq:unpjeth}, which arises in the azimuthal distribution of unpolarized hadrons in the unpolarized electron and proton collisions. At the partonic level, the relevant structure function, $F_{UU,U}^{\cos(\phi_{q}-\hat{\phi}_h)}$, arises  from the Boer-Mulders function $h_1^\perp$ in the unpolarized proton coupled with the Collins fragmentation function $H_1^\perp$ for the final-state hadron inside the jet. 

To begin with, we define the new azimuthal asymmetry $A_{UU,U}^{\cos(\phi_{q}-\hat{\phi}_h)}$ by normalizing the structure function $F_{UU,U}^{\cos(\phi_{q}-\hat{\phi}_h)}$ by the unpolarized and azimuthal-independent structure function $F_{UU,U}$, $A_{UU,U}^{\cos(\phi_{q}-\hat{\phi}_h)}=F_{UU,U}^{\cos(\phi_{q}-\hat{\phi}_h)}/F_{UU,U}$. Here, the denominator $F_{UU,U}$ and the numerator $F_{UU,U}^{\cos(\phi_{q}-\hat{\phi}_h)}$ can be factorized as follows
\bea
\label{eq:FUUUbefore}
F_{UU,U} =&\hat{\sigma}_0\,H(Q)\sum_qe_q^2\, {\mathcal{D}}_{1}^{h/q}(z_h,j_\perp^2)\int\frac{b \,db}{2\pi}J_0(q_Tb)\,x\,\tilde{f}^{q}_{1}(x,b^2)\bar{S}(b^2,R)\,,\\
F_{UU,U}^{\cos(\phi_{q}-\hat{\phi}_h)} =&\hat{\sigma}_T\,H(Q)\sum_q e_q^2\, \frac{ j_\perp}{z_hM_h} M \mathcal{H}_1^{\perp\,h/q}(z_h,j_\perp^2)\int\frac{b^2 \,db}{2\pi}J_1(q_Tb)\,x\,\tilde{h}_1^{\perp\, q(1)}(x,b^2)\bar{S}(b^2,R)\,,
\label{eq:FUUU-spin}
\eea
where $\hat{\sigma}_0=\frac{\alpha_{\rm em}\alpha_s}{sQ^2}\frac{2(\hat{u}^2+\hat{s}^2)}{\hat{t}^2}$, $\hat{\sigma}_T=\frac{\alpha_{\rm em}\alpha_s}{sQ^2}\left(\frac{-4\hat{u}\hat{s}}{\hat{t}^2}\right)$, and the soft function $\bar{S}(b^2,R)=\bar{S}_{\rm global}(b^2)\bar{S}_{cs}(b^2,R)$, with the global soft function $\bar{S}_{\rm global}$ and $\bar{S}_{cs}$ given in~\cite{Kang:2021ffh}.

\begin{figure}
\centering
\includegraphics[width = 0.31\textwidth]{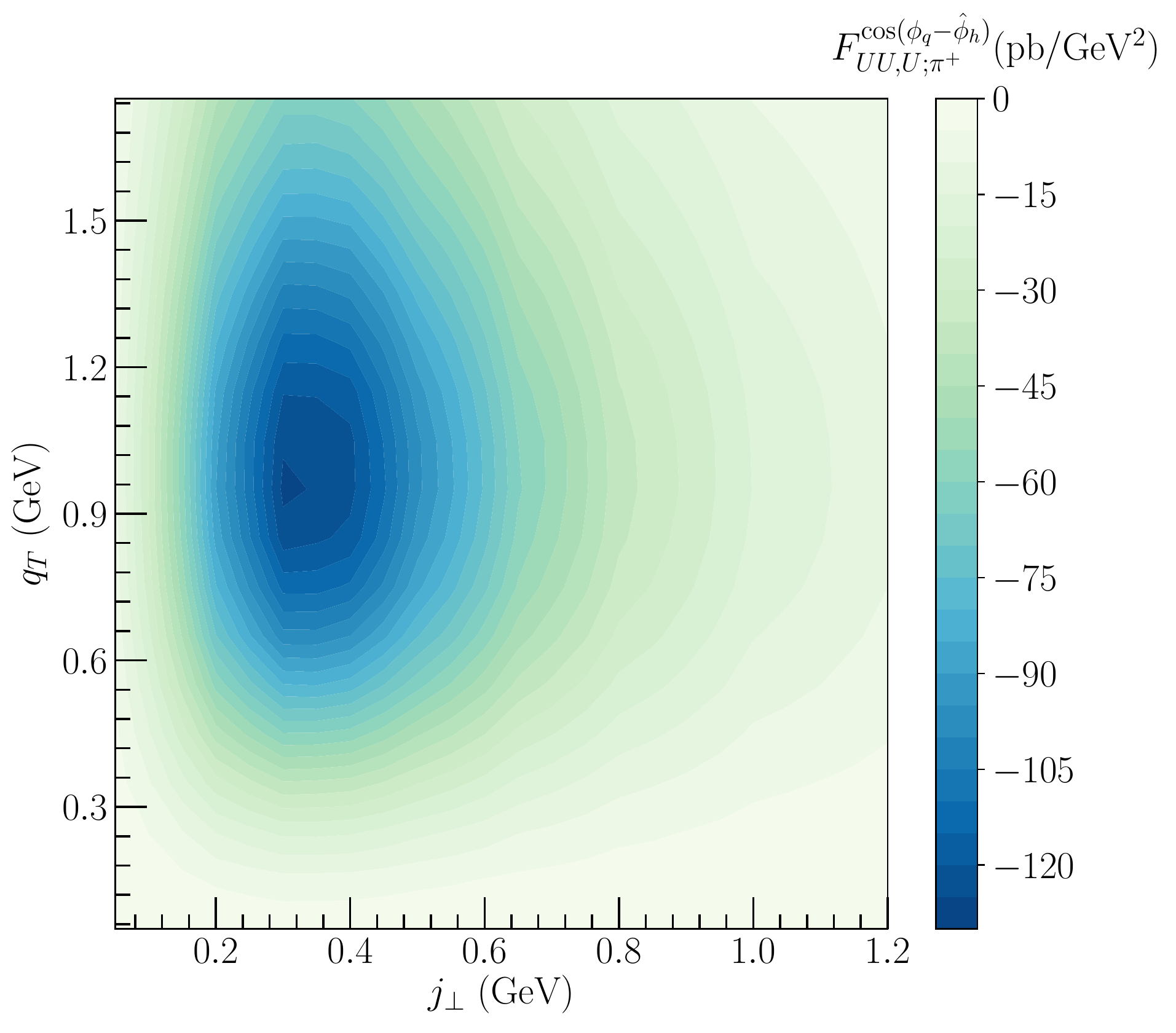}
\includegraphics[width = 0.30\textwidth]{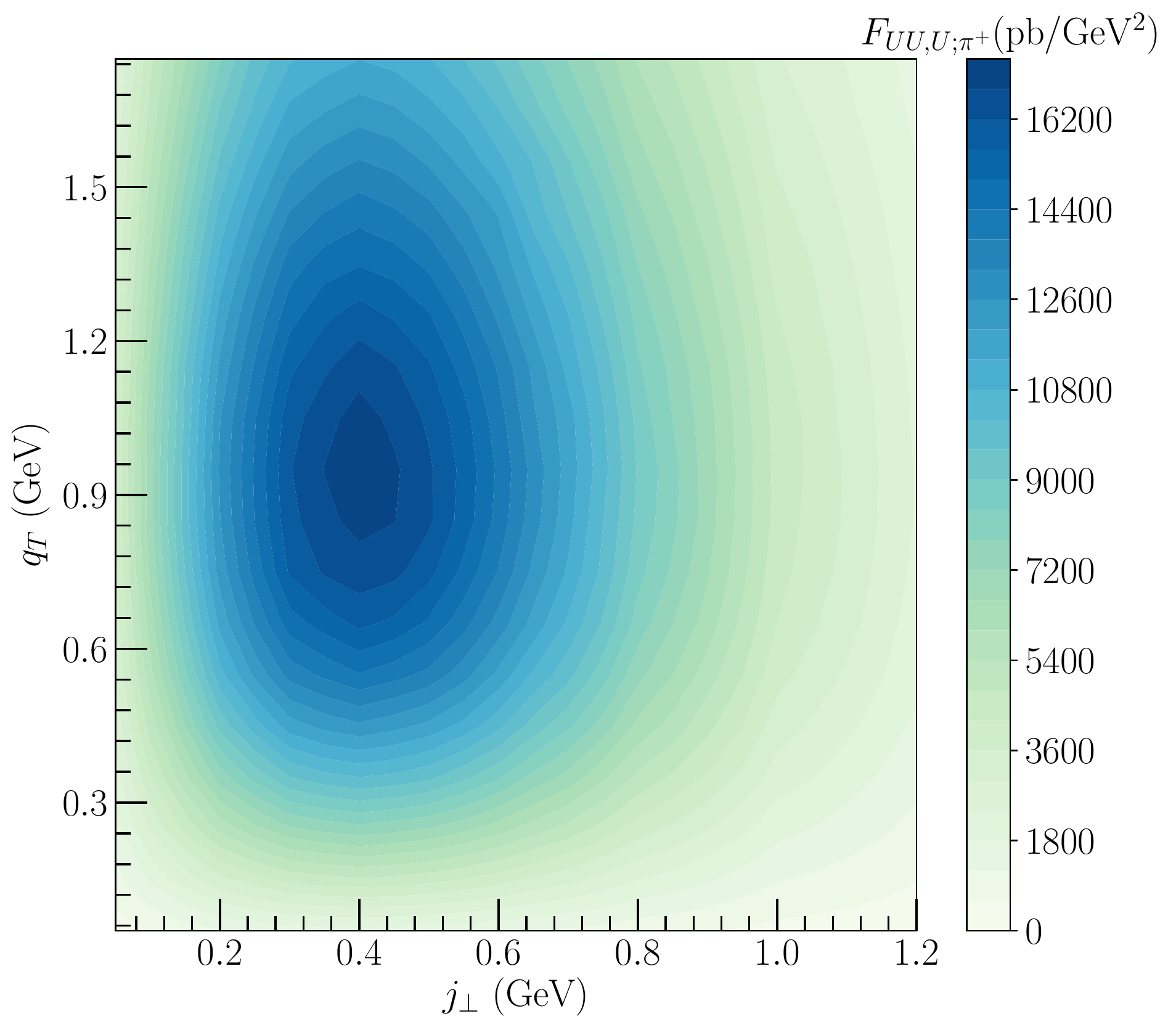}
\includegraphics[width = 0.31\textwidth]{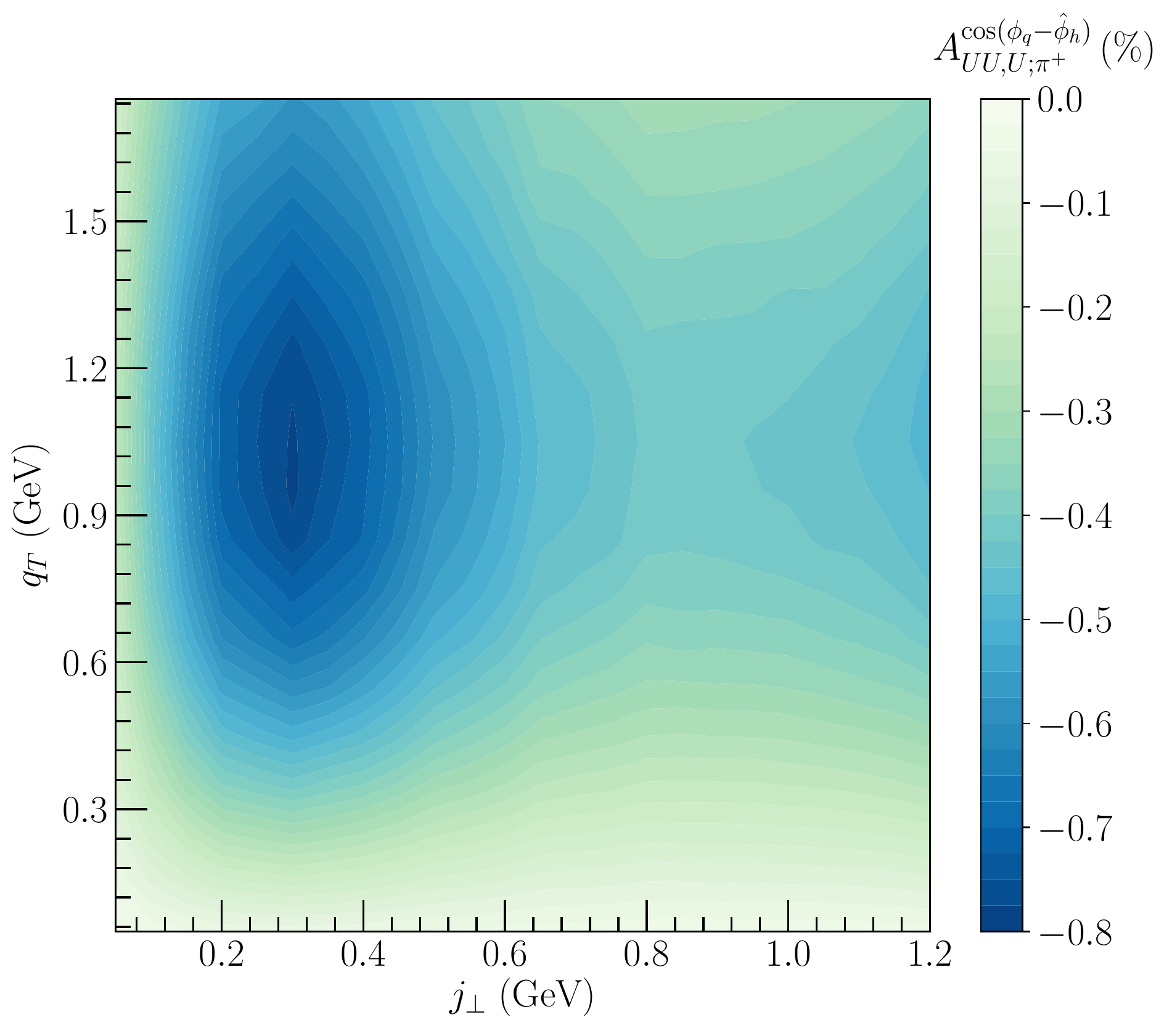}\\
\includegraphics[width = 0.31\textwidth]{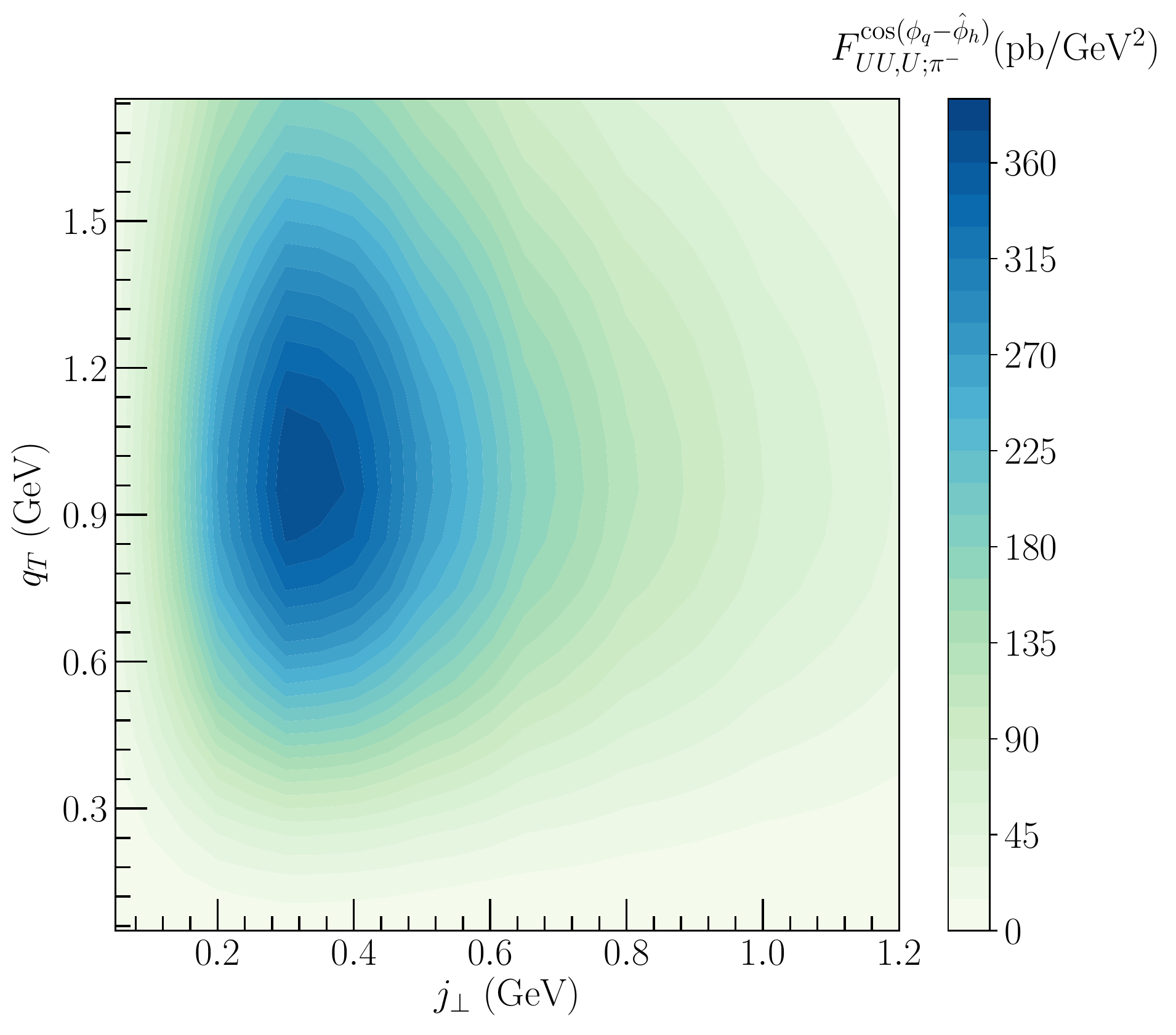}
\includegraphics[width = 0.30\textwidth]{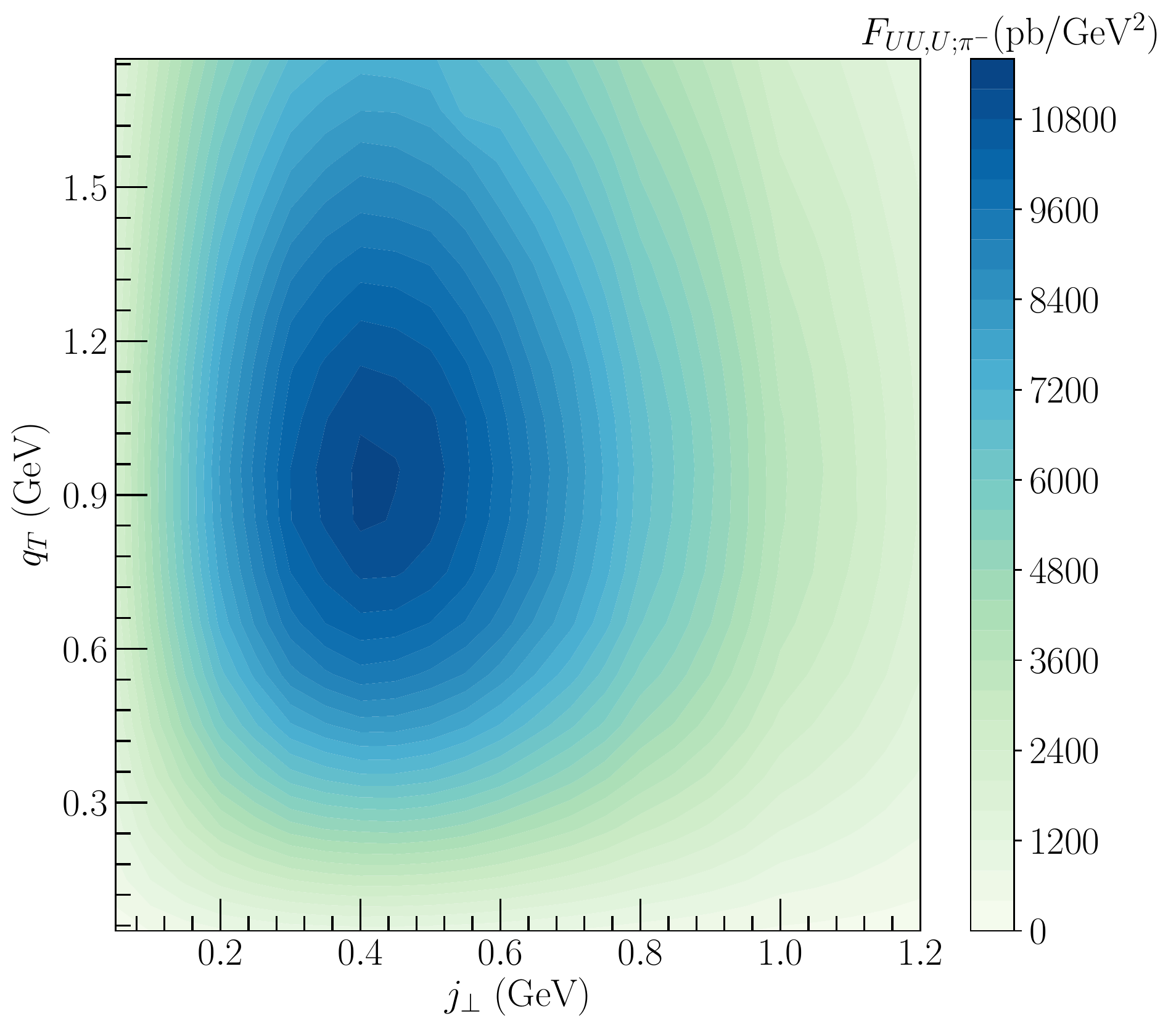}
\includegraphics[width = 0.31\textwidth]{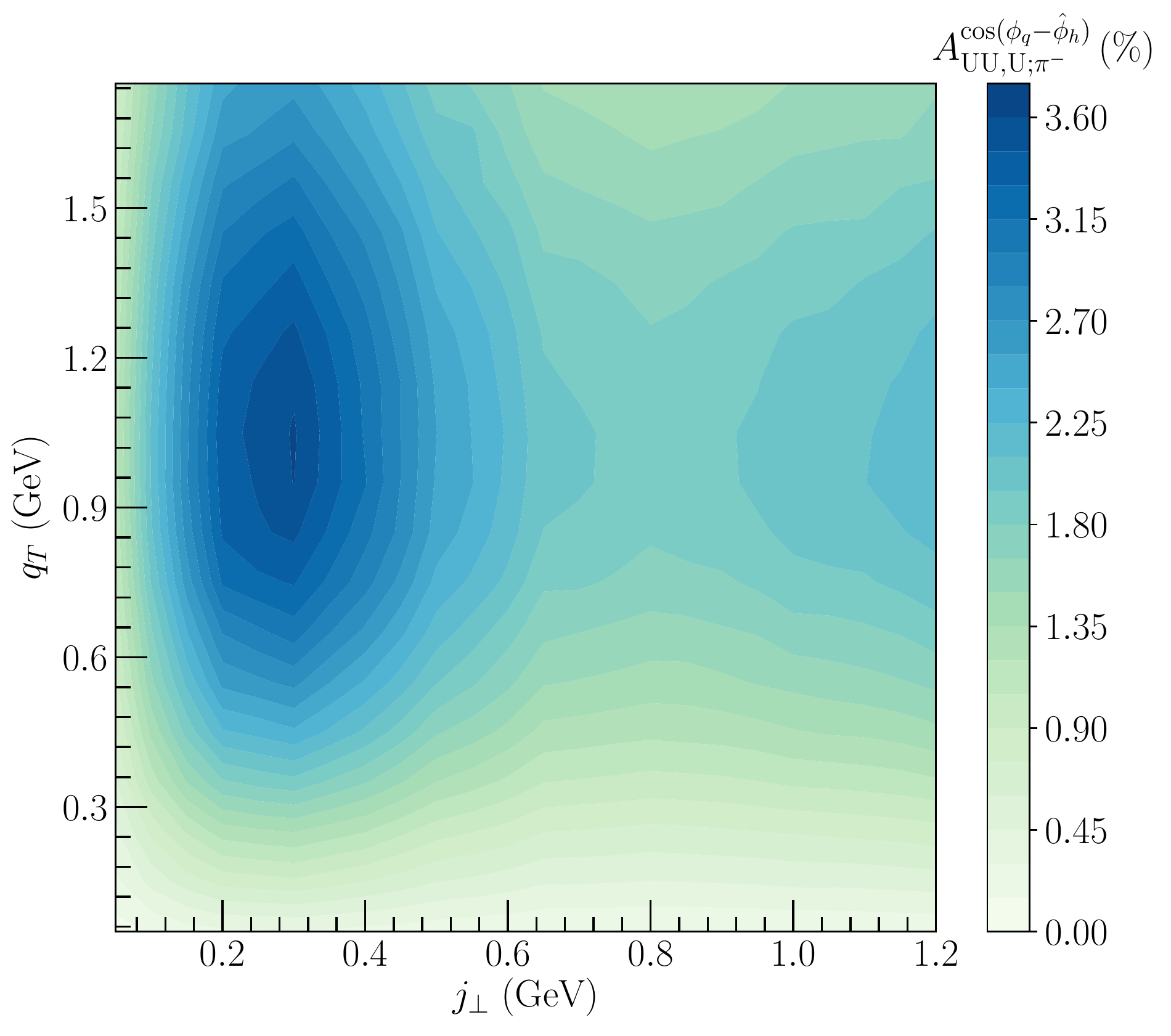}
\caption{The contour plots of $F_{UU,U}^{\cos(\phi_{q}-\hat{\phi}_{h})}$ (left column), $F_{UU,U}$ (middle column) and their ratio $A_{UU,U}^{\cos(\phi_{q}-\hat{\phi}_{h})}$ (right column) as a function of $q_T$ and $j_\perp$ for unpolarized $\pi^+$ (upper panel) and $\pi^-$ (lower panel) in jet production.}
  \label{fig:scn2}
\end{figure}

\begin{figure}
\centering
\includegraphics[width = 0.29\textwidth]{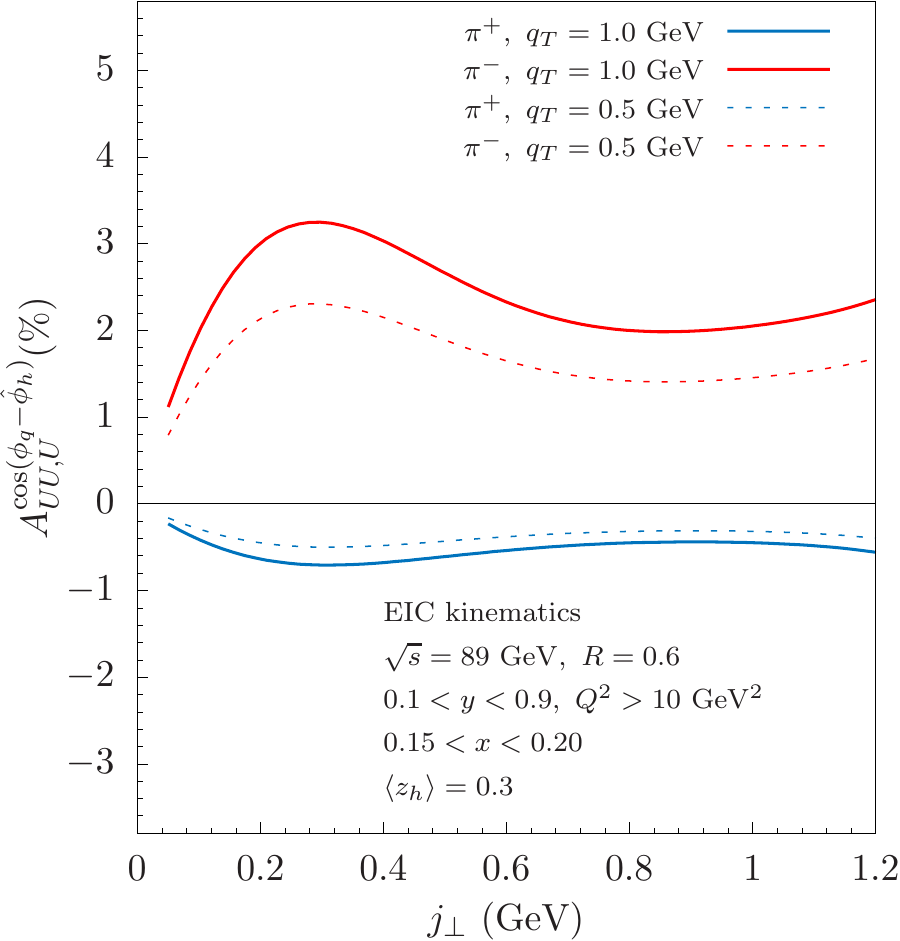}
\hspace{1cm}\includegraphics[width = 0.29\textwidth]{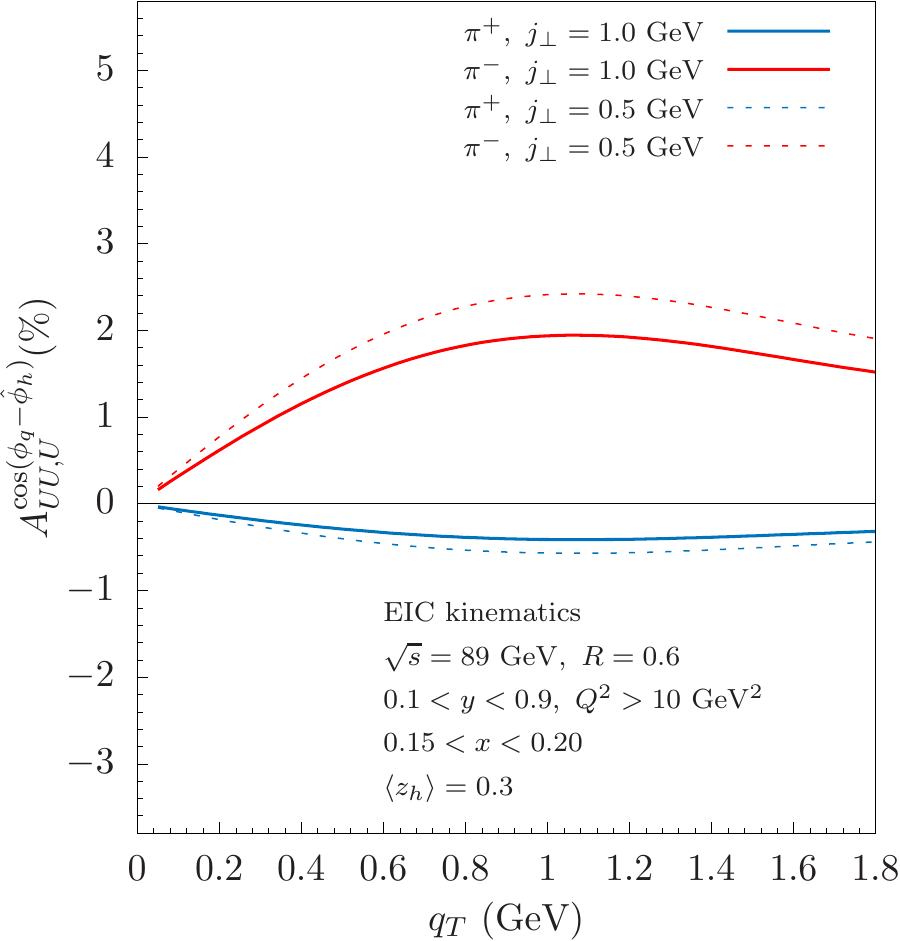}
\caption{Horizontal slices ($j_\perp$-dependent, left plot) and vertical slices ($q_T$-dependent, right plot) of $A_{UU,U}^{\cos(\phi_{q}-\hat{\phi}_{h})}$ in Fig.~\ref{fig:scn2} for unpolarized $\pi^\pm$ in jet production with electron in unpolarized $ep$ collision at the EIC kinematics.}
  \label{fig:scn2_2d}
\end{figure}

For the phenomenological study, we emply the Boer-Mulder functions extracted in~\cite{Barone:2009hw} and the Collins TMDFFs extracted in~\cite{Kang:2015msa}. In Fig.~\ref{fig:scn2}, we plot the structure functions $F_{UU,U}^{\cos(\phi_{q}-\hat{\phi}_{h})}$, $F_{UU,U}$ and asymmetry $A_{UU,U}^{\cos(\phi_{q}-\hat{\phi}_{h})}$ differentiated in $q_T$ and $j_\perp$ with $0.15<x<0.2$, jet radius $R=0.6$, inelasticity cut $0.1<y<0.9$, momentum fraction $\langle z_h\rangle=0.3$ and $Q^2>10$ GeV$^2$ for $\pi^+$ (upper panel) and $\pi^-$ (lower panel) production inside the jet. In the first column, the $q_T$-dependence is determined by the Boer-Mulders function and the $j_\perp$-dependence is determined by the Collins function. As for the second column, the Sudakov peak is featured by the unpolarized TMDPDF and TMDFF for constant $j_\perp$ and $q_T$ slices. Finally, in the asymmetry plot (the third column), the spin asymmetry tends to be negative $\sim 1\%$ for $\pi^+$ production in the jet and positive with magnitude $\sim3\%$ for $\pi^-$ production in the jet. 
To better illustrate the results, in Fig.~\ref{fig:scn2_2d} we present the horizontal ($j_\perp$-dependent) slices for $q_T=1.0$ GeV (solid curves) and $q_T=0.5$ GeV (dashed curves) of the third column of Fig.~\ref{fig:scn2} for $\pi^+$ (blue curves) and $\pi^-$ (red curves) productions inside a jet in the left plot. As for the right plot of Fig.~\ref{fig:scn2_2d}, we show the vertical ($q_T$-dependent) slices for $j_\perp=1.0$ GeV (solid curves) and $j_\perp=0.5$ GeV (dashed curves) of the $A_{UU,U}^{\cos(\phi_{q}-\hat{\phi}_{h})}$ plot. With a measurable asymmetry of a few percents for $\pi^\pm$ with the TMD evolution, this asymmetry can serve as a potential observable at the EIC for probing the Boer-Mulders functions and Collins fragmentation functions.
\section{Example 2: Transversely polarized $\Lambda$ in jet}\label{sec:pol_h}
In this section, we predict for the transverse spin asymmetry of a $\Lambda$ particle inside the jet as an example of studying the back-to-back $e$-jet production with a polarized hadron in jet. The azimuthal asymmetry $A^{\sin(\hat{\phi}_\Lambda-\hat{\phi}_{S_\Lambda})}_{UU,T}$ arises from the structure function $F^{\sin(\hat{\phi}_\Lambda-\hat{\phi}_{S_\Lambda})}_{UU,T}$ divided by $F_{UU,U}$, where the structure function $F_{UU,U}$ has been introduced in Eq.~\eqref{eq:FUUUbefore} and $F^{\sin(\hat{\phi}_\Lambda-\hat{\phi}_{S_\Lambda})}_{UU,T}$ is factorized as 
\bea
\label{eq:FUUT-lambda}
F^{\sin(\hat{\phi}_\Lambda-\hat{\phi}_{S_\Lambda})}_{UU,T} =&\hat{\sigma}_0 \,H(Q)\sum_q e_q^2\, \frac{ j_\perp}{z_hM_h} \mathcal{D}_{1T}^{\perp\,\Lambda/q}(z_\Lambda,j_\perp^2)\int\frac{b \,db}{2\pi}J_0(q_Tb)\,x\,\tilde{f}_1^{q}(x,b^2)\bar{S}(b^2,R)\,,
\eea
where the polarizing TMDJFF $\mathcal{D}_{1T}^{\perp\,\Lambda/q}$ can be connected to the polarizing TMDFF $D_{1T}^{\perp\,\Lambda/q}$ \cite{Kang:2020xyq}. In our numerical analysis, we provide results for $\Lambda$ transverse polarization inside the jet with EIC kinematics by considering TMD evolution in both $D_1^{\Lambda/q}$ and $D_{1T}^{\perp\,\Lambda/q}$. We have applied the AKK08 parametrizations~\cite{Albino:2008fy} for the collinear FF $D_1^{\Lambda/q}(z_\Lambda, \mu_{b_*})$ and extraction in~\cite{Callos:2020qtu} for $D_{1T}^{\perp\,\Lambda/q}$.
\begin{figure}[t]
\centering
\includegraphics[width = 0.33\textwidth]{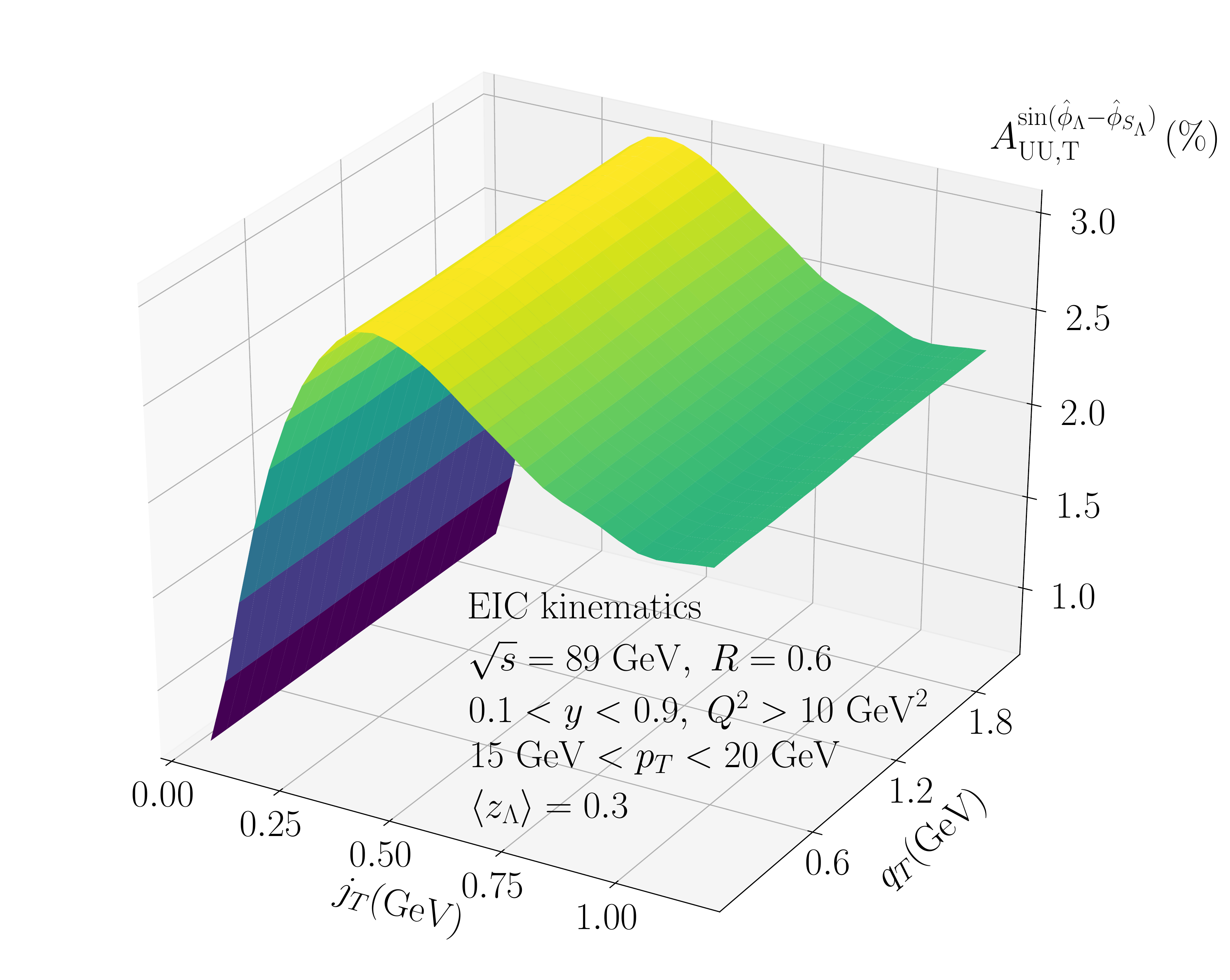}
  \includegraphics[width = 0.315\textwidth]{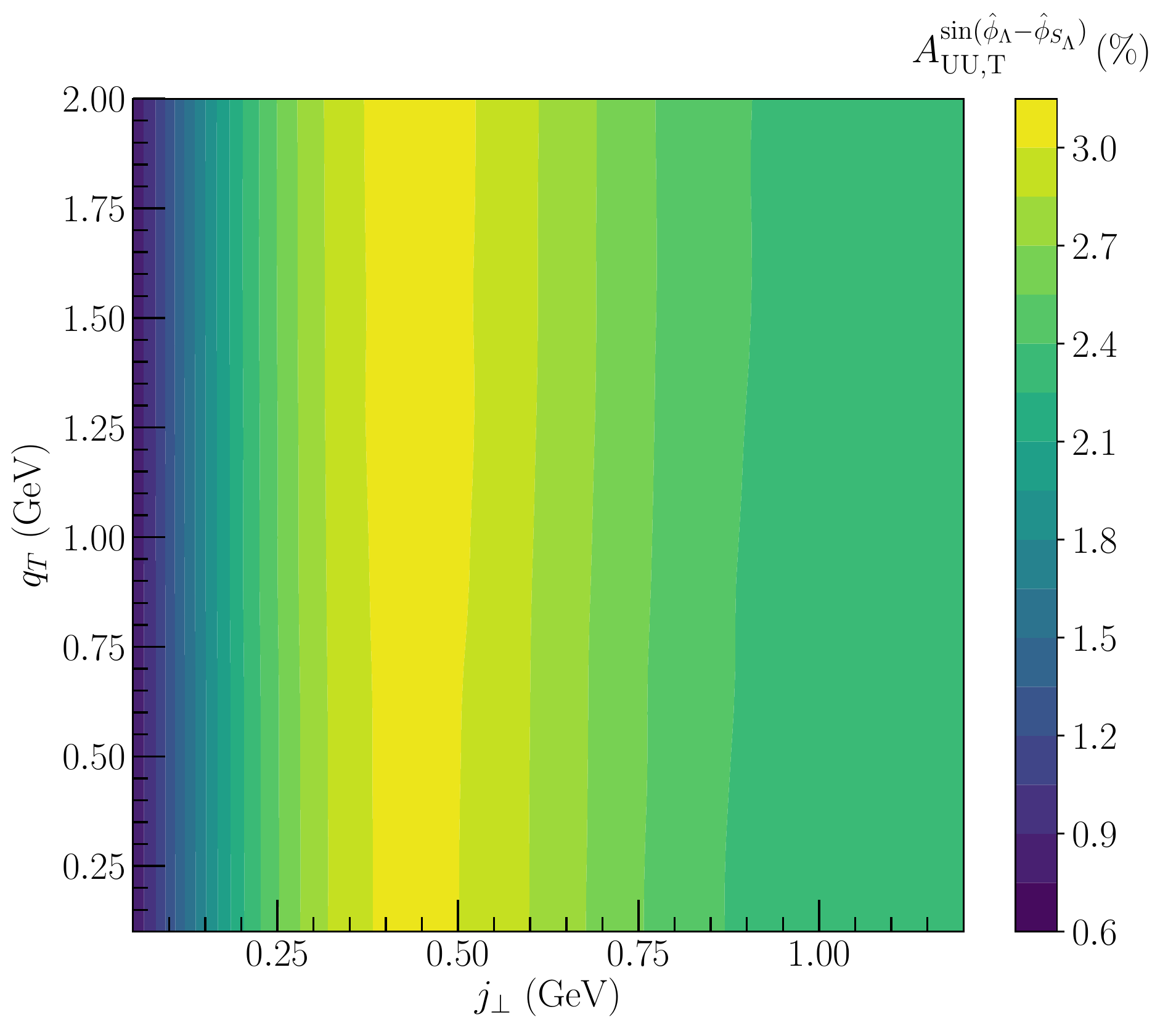}
  \includegraphics[width = 0.285\textwidth]{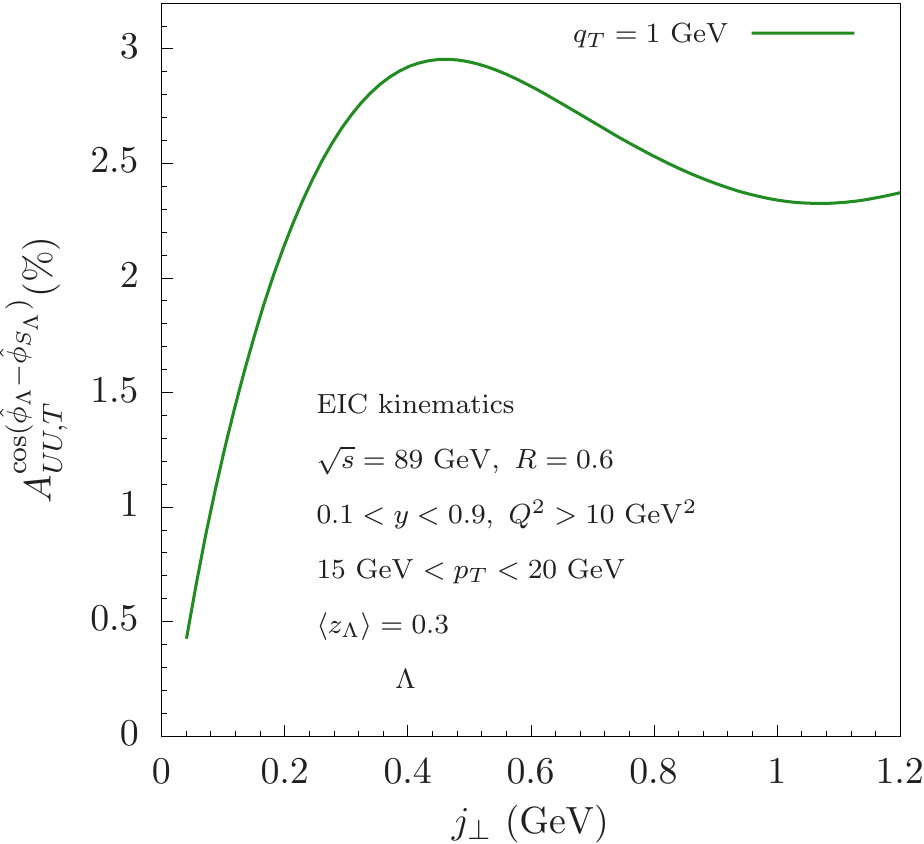}
\caption{$A_{UU,T}^{\sin(\hat{\phi}_{\Lambda}-\hat{\phi}_{S_\Lambda})}$ differentiated in the imbalance $q_T$ and momentum $j_\perp$ for transversely polarized $\Lambda$ in jet production. Left: three-dimensional plot. Middle: contour plot. Right: Horizontal slices of the contour plot.}
  \label{fig:scn3}
\end{figure}
Next, we plot the asymmetry $A_{UU,T}^{\sin(\hat{\phi}_{\Lambda}-\hat{\phi}_{\Lambda})}$ as a function of the momentum imbalance $q_T$ and transverse momentum $j_\perp$ using $\sqrt{s}=89$ GeV with $0.1<y<0.9$ and $0.15<x<0.20$ in Fig.~\ref{fig:scn3}. Since the unpolarized TMDPDF $f_1$ determines both the $q_T$ dependence of the numerator and denominator of the asymmetry,  we discover that in constant $j_\perp$ slices, the asymmetry is constant as expected. Due to the cancellation of the dependence in TMDPDFs in the ratio, $A_{UU,T}^{\sin(\hat{\phi}_{\Lambda}-\hat{\phi}_{\Lambda})}$ is especially effective for extracting $D_{1T}^{\perp\,\Lambda/q}$. Then in the right panel of Fig.~\ref{fig:scn3}, we show horizontal slices of the contour plots as a function of $j_\perp$ for the EIC kinematics, where the asymmetry increases up to $3\%$ at $j_\perp=0.4$ GeV and then gradually drops to around $2.5\%$, indicating practicability for measurements at the future EIC.

\section{Summary}\label{sec:summary}
In this work, we investigate all possible azimuthal asymmetries that can arise in back-to-back electron-jet production with the hadron observed inside the jet in electron-proton collisions. We highlight the strengths of such observables in constraining TMDPDFs and TMDFFs separately by measuring simultaneously the electron-jet transverse momentum imbalance $\bm{q}_T$ (with respect to the beam direction) and the hadron transverse momentum $\bm{j}_\perp$ with respect to the jet axis. We show two azimuthal/spin asymmetries for hadron inside the jet in the phenomenological studies with EIC kinematics. We demonstrate that the electron-jet production is an excellent way for studying transverse momentum-dependent functions.

\end{document}